\documentclass{PoS}
\usepackage{amsmath,amssymb,verbatim}
\def\Sw{S}
\def\Sc{S_{\rm cl}}

\def\s0#1#2{\mbox{\small{$ \frac{#1}{#2} $}}}
\def\0#1#2{\frac{#1}{#2}}
\def\eq#1{(\ref{#1})}

\def\n{\nabla}
\def\tn{\hat{\nabla}}

\def\vp{\varphi}

\def\Sdet{{\rm SDet}}
\def\Str{{\rm STr}}
\def\M{M}
\newcommand{\Mc}{\tilde{M}}

\newcommand{\al}{\alpha}
\newcommand{\alm}{\alpha^{-1}}

\def\half{\frac{1}{2}}
\def\veps{\varepsilon}
\def\bpsi{\bar{\psi}}
\def\bveps{\bar{\varepsilon}}
\newlength{\hack}
\newcommand{\sectionn}[1]{\vspace{-\hack}\section{#1} \vspace{-\hack}}
\title{Generalisations of the Ginsparg-Wilson relation and a remnant
  of supersymmetry on the lattice}

\ShortTitle{A  remnant of supersymmetry on the lattice}

\author{\speaker{Georg Bergner}
\\
        Theoretisch-Physikalisches Institut, Universit\"at Jena,
  D-07743 Jena, Germany\\
        E-mail: \email{g.bergner@tpi.uni-jena.de}}

\author{Falk Bruckmann\\
 Institut f\"ur Theoretische Physik, Universit\"at
  Regensburg, D-93040 Regensburg, Germany\\
        E-mail: \email{falk.bruckmann@physik.uni-regensburg.de}}

\author{Jan M.~Pawlowski\\
 Institut f\"ur Theoretische Physik, Universit\"at
  Heidelberg, D-69120 Heidelberg, Germany\\
        E-mail: \email{j.pawlowski@thphys.uni-heidelberg.de}}

      \abstract{We introduce a lattice symmetry relation for field
        theories with general linear symmetries. For chiral symmetry
        the well-known Ginsparg-Wilson relation is reproduced. The new
        relation encodes the remnant of the original symmetry on the
        lattice and guides the construction of invariant lattice
        actions. We apply this approach to lattice supersymmetry.
        There, an additional constraint has to be satisfied which
        originates in the derivative operator in the symmetry
        transformations. As a consequence the non-local SLAC
        derivative operator appears in the lattice transformation. Despite this
        non-local form we show how local solutions for quadratic
        actions can be found. For interacting theories the relation in
        general leads to a non-polynomial action that can be reduced
        to a finite polynomial order only under certain conditions.}

\FullConference{The XXVI International Symposium on Lattice Field Theory\\
		 July 14-19 2008\\
		 Williamsburg, Virginia, USA}

\begin{document}


\sectionn{Introduction}
One of the basic problems of lattice simulations of quantum field
theory is the continuum limit.  The specific form of the lattice
action governs not only the renormalisation; in practice it also
decides, whether it is possible to approach the continuum theory
sufficiently fast for the numerical resources at hand.

To preserve or recover the symmetries of the continuum theory is an
essential part of the continuum limit. In some important cases
continuum symmetries are necessarily broken on the lattice. This
applies in particular to space-time symmetries, as the lattice
introduces a hard breaking of the rotational and translational
invariance of the continuum.  Supersymmetry is another example for
this situation that indeed relates to the aforementioned space-time
symmetry. On the technical level this manifests itself in the failure
of the Leibniz rule on the lattice \cite{Dondi:1976tx}, which makes it
impossible to find a local lattice action preserving the full
supersymmetry of an interacting theory.  With no realisation of
supersymmetry at finite lattices, the continuum limit is thus very
hard to control.

Related problems are the representation of the continuum anomalies on
the lattice and the different possible realisations of symmetry
operators on the lattice, which can not all be expected to lead to the
correct continuum limit.  The naive realisation of chiral symmetry on
the lattice, for example, can only be achieved at the cost of an
unwanted doubling of the degrees of freedom, as stated by the
Nielsen-Ninomiya theorem \cite{Nielsen}. For symmetries broken on the
lattice, the symmetric continuum theory may be achieved by fine-tuning
the parameters of the action. This type of renormalisation procedure,
however, is numerically rather expensive and requires knowledge of
physical quantities that should actually be a result of the numerical
calculations.

The approach of Ginsparg and Wilson \cite{Ginsparg:1981bj} to chiral
symmetry resolves this fine-tuning problem within a renormalisation
group setting.  The procedure is based on a block spin
transformation, i.e.\ a mapping from the continuum to the lattice (or
from a finer lattice to a coarser one).  This guarantees the minimal
amount of symmetry breaking at each blocking step and also ensures a
definite interpretation of the lattice observables in terms of
continuum quantities and the correct continuum limit. In principle
this setting can also be used to calculate an effective lattice
action, but explicit solutions are calculable only in very special
cases. It is, however, possible to obtain the implications of the
continuum symmetries purely on the level of the effective lattice
action: continuum symmetries are translated into lattice relations
that can be interpreted as deformed lattice symmetries. Lattice
implementations which obey these relations ensure the correct
symmetries in the continuum limit. Moreover, the deformed symmetry
operators provides us with the correct realisation of the continuum
symmetry on the lattice.  Such an approach is closely related to the
renormalisation group studies in the continuum. The lattice is in that
case just a special regulator of the theory and the lattice symmetry
relations correspond to the quantum master equations or modified
Slavnov-Taylor identities,
for a review see \cite{Pawlowski:2005xe}. 

Here we consider the blocking procedure for general continuum theories
with general linear symmetries \cite{Bergner:2008ws}. This systematic
approach enables us to derive a relation for lattice actions with
remnant lattice (super)symmetry. Locality problems of the naive
implementation of (super)symmetry might be cured by an appropriately
chosen blocking term leading to a deformed lattice (super)symmetry. We
discuss both quadratic theories, revisiting the chiral case as a
pedagogical example, as well as interacting theories.  Evidently, the
main obstructions for the solutions of the symmetry relation will be
space-time locality and a polynomial nature in the fields.

\sectionn{Blocking transformation}

The blocking transformation specifies the map from the continuum
theory onto the lattice. In particular this map entails
how lattice symmetry transformations derive from their continuum
analogues, and which symmetry relation the blocked lattice action has
to satisfy.

The first step of the blocking procedure is an averaging of the
continuum fields $\varphi(x)$ around each lattice point $n$, done with
an averaging function $f(x)$,
\begin{equation}
 \Phi_{n}[\varphi]:=\int\!\!\!dx\; f(x-x_n)\varphi(x)\;,
\label{eq:step1}\end{equation} 
that is typically peaked around zero. In this way we have defined
averaged fields $\Phi_n$ with an integer lattice index $n$.  This
averaging can easily be extended to more than one dimension.

The second step introduces a blocking matrix $\alpha$ that connects
the averaged fields $\Phi_n$ to the lattice fields $\phi_n$ via a
Gaussian factor,
\begin{equation}
\label{eq:wilson_action}
e^{-S[\phi]}:=\frac{1}{\Sdet^{1/2}\,\alpha}\int\!\!\!d\vp\; 
e^{-\half(\phi-\Phi[\vp])_n\al_{nm}(\phi-\Phi[\vp])_m}
\;e^{-\Sc[\vp]}\;,
\end{equation}
which can be read as smearing in field space. It defines a blocked
lattice action $S[\phi]$ for a given continuum action $S_{\rm
  cl}[\vp]$ and blocking kernel $\al$.

This prescription has a simple interpretation if $f(x-x_n)$ approaches
$\delta(x-x_n)$ and the eigenvalues of $\alpha$ diverge in the
continuum limit: then the lattice action becomes the continuum action.
The first demand is actually a necessary one in order to approach the
continuum, since $f$ governs the resolution of the lattice.  For more
general $\alpha$'s one has to investigate the generating functional
with the action $S[\phi]$ defined above, because this quantity
determines the meaning of the observables in the $\phi$ theory.
Performing Gaussian integrations one gets an exact agreement,
\begin{equation}
  \int\!\!d\phi\; e^{-S[\phi]+J\phi}=e^{-J\alm J}\int\!\!d\vp\; 
  e^{-\Sc[\vp]+J\Phi[\vp]} \;,
\end{equation}
up to the $J$-dependent prefactor that can be easily calculated
knowing $\alpha$.

The two steps explained above will be investigated w.r.t.\ their
effect on lattice symmetries in the next two sections. We shall see
that for global symmetries such as chiral symmetry the first step
is trivial.  For a general space-time-dependent symmetry, and
especially supersymmetry, also the first step is non-trivial and leads
to obstructions. 

\sectionn{Lattice symmetry: the blocked generator \ldots}
Here we consider a continuum action $\Sc[\vp]$ that is invariant under
an infinitesimal continuum variation $\tilde{\delta}\vp^i=
\veps\Mc^{ij}\vp^j$. $\veps$ is an infinitesimal parameter, bosonic or
fermionic, and the superscript $i$ labels internal indices and species
of fields. For example, in the case of chiral symmetry $\varphi^i$ are
the components of a spinor, and $\tilde M$ is a (global)
chiral rotation.

After the averaging step \eq{eq:step1} this symmetry must act only
among the discrete fields labelled by $n$. This means that the
transformation must be `pulled outside' of the averaging procedure as
\begin{equation}
\label{eq:addconst}
\int\!\!\!dx\; f(x-x_n)\; \veps{\Mc^{ij}}\vp^j(x)=
\veps\M^{ij}_{nm}\Phi_m[\vp^j]=\veps{\M_{nm}^{ij}}
\int\!\!\!dx\; f(x-x_m)\varphi^j(x)\quad \forall\varphi^j(x)\; ,
\end{equation}
to arrive at a lattice transformation $M$ and the corresponding variation
$\delta\phi^i_n= \veps\M^{ij}_{nm}\phi^j_m$.

It is trivial to compute $\M$ for a symmetry that acts only on
multiplet indices and \eq{eq:addconst} was therefore not taken into
account in the discussions of chiral symmetry.  Starting with a
continuum transformation this procedure defines a corresponding
lattice counterpart.  Note, however, that for space-time dependent
symmetries one cannot find such an $\M$ for every $\Mc$ and $f$. Then
\eq{eq:addconst} constitutes an additional constraint for $\Mc,M$ and
$f$ and is not a mere definition of $\M$.  This applies in particular
to the derivative operator in the continuum supersymmetry $\Mc$, a
problem\footnote{This constraint has been mentioned without further
  investigation in \cite{So:1998ya}.} to which we will come back in
section \ref{sect:add}.

We conclude that an invariance under the naive symmetry
transformations $\delta\phi$ is the first guess for the resulting
lattice symmetry. However, in order to calculate the full effect of
the blocking on the symmetry one must also include the blocking matrix
$\alpha$ as done in the next section.
\sectionn{\ldots and the symmetry
  relation}
The symmetry relation for the lattice action is derived from applying
an infinitesimal transformation on the lattice fields, $\phi$ in
\eq{eq:wilson_action}.  This transformation can be absorbed within a
symmetry transformation of the continuum fields $\varphi$ on the rhs.\
of \eq{eq:wilson_action} using the additional constraint. More details
of the calculation can be found in \cite{Bergner:2008ws}. We finally 
arrive at the following relation
\begin{equation}
\label{eq:relation}
 \M_{nm}^{ij}\phi_m^j \0{\delta \Sw
}{\delta \phi_n^i} =
(\M \alpha^{-1})_{nm}^{ij} \left( \0{ \delta
\Sw}{\delta \phi_m^j}\0{\delta \Sw }{\delta \phi_n^i}- \0{ \delta^2
\Sw}{\delta \phi_m^j\delta \phi_n^i}\right)
+\big(\Str
\M-\Str\Mc\big)\; .
\end{equation}
The trace part $\Str\, \Mc$ takes care of an infinitesimal change of
the measure $d\vp$, hence the last term represents the difference 
between continuum and lattice anomalies.

The relation \eq{eq:relation} entails the deformation of the continuum
symmetry on the lattice due to the chosen blocking kernel $\alm$. For
symmetric blocking kernels the rhs.\ of \eq{eq:relation} vanishes and
the lattice theory respects the symmetry generated by $M$, $\delta S=\veps
M\phi\delta S/\delta\phi=0$. 
More specifically, 
a symmetric 
part $\alm_S$ does not contribute to \eq{eq:relation} if 
it satisfies
\vspace*{-0.2cm}
\begin{equation}
\label{eq:symmetric_alpha}
\M\alpha^{-1}_S\pm (\M\alpha^{-1}_S)^T=0 
\vspace*{-0.2cm}
\end{equation}
(the minus sign applies when the symmetry transforms fermions into
fermions). 
It is therefore the symmetry breaking part of the blocking
kernel that is responsible for the difference between the naive and
the deformed lattice symmetry, by generating non-linear terms on the
rhs.\ of \eq{eq:relation}.  We emphasise that the averaging function
$f$, which is a key ingredient of the mapping onto the lattice,
appears in this relation only in terms of the naive lattice symmetry
operator~$M$.
\sectionn{Quadratic lattice actions and chiral symmetry}
\label{sec:quadact}
To achieve a better understanding of the above relation
(\ref{eq:relation}), we discuss it in the simplest case of quadratic
actions. Although this seems to be a rather trivial case, it already
includes the well known Ginsparg-Wilson (GW) relation for chiral symmetry.
There the Dirac action, relevant for this symmetry, is
quadratic for a given gauge background. 

Evaluating our relation (\ref{eq:relation}) for a quadratic lattice
action, $S=\half \phi_n^i K_{nm}^{ij}\phi_{m}^{j}$, we obtain 
a matrix equation from the terms  quadratic in the fields 
(for the field-independent trace part see \cite{Bergner:2008ws}),
\begin{equation}
\label{eq:matrix}
\M^T K \pm (\M^T K)^T = K^T (\M \alm)^T K\pm (K^T (\M \alm)^T K)^T\; .
\end{equation}
Again, the minus sign applies when the symmetry transforms fermions into
fermions.  

Now we discuss the specific case of chiral symmetry with symmetry
operator $\Mc,M\sim \gamma_5$ and the simplest blocking kernel
$\alpha_{nm}=\frac{1}{a}\delta_{nm}$, that is diagonal in lattice
sites and indeed diverging in the continuum limit. Then the matrix
relation \eq{eq:matrix} is nothing but the Ginsparg-Wilson relation
$\{\gamma_5,D\}=a D \gamma_5 D$ \cite{Ginsparg:1981bj}. Furthermore
\eq{eq:matrix} can be rewritten as
\begin{equation}\label{eq:symm}
\M_{\rm def}^T K\pm (\M_{\rm
def}^T K)^T=0\;, \quad \text{with} 
\quad M_{\rm def}=\M(1-\alm K)\; .
\end{equation}
Eq.~(\ref{eq:symm}) entails the invariance of the action under a
transformation generated by $M_{\rm def}$.  In other words, the
invariance of the continuum action always implies an invariance of the
lattice action.  There are, however, other conditions for $M_{\rm
  def}$ to qualify as a lattice symmetry: Clearly $M_{\rm def}$ has to
approach its continuum counterpart in the continuum limit.
Furthermore, $M_{\rm def}$ must be local to define a proper lattice
symmetry, in order to have lattice artefacts of the symmetry under
control in the continuum limit. For non-local lattice artefacts the
continuum limit is at stake. We conclude that the freedom to choose a
blocking kernel must be utilised to find a local deformed symmetry.
We will denote only invariances satisfying the conditions explained
above as lattice symmetry. In turn, non-local invariances should not 
be seen as lattice symmetries. 

In the case of chiral symmetry with blocking kernel
$\alpha_{nm}=\frac{1}{a}\delta_{nm}$ one arrives at deformed
$\gamma_5$'s, now depending on the Dirac operator and hence the
background field, cf.\ \cite{Luscher}. The locality conditions
excludes for instance Wilson fermions as a solution for the chiral
symmetry. Indeed, locality is the main issue about solutions of the
Ginsparg-Wilson relation. This problem already occurs for quadratic
theories and we will face it again for supersymmetry below.
\sectionn{A derivative operator for lattice supersymmetry}
\label{sect:add}
Now we return to the additional constraint (\ref{eq:addconst}) with
$\Mc$ containing the derivative operator, as is the case for
supersymmetry. The constraint entails, that the averaged derivative of
any continuum field must be represented by a linear combination of
averaged fields living on the lattice sites only. The corresponding
coefficients constitute the lattice derivative operator $\n$. On the
averaging function $f$ itself and transformed into Fourier-space the
constraint reads
\begin{equation}
  [\n(p_k)-ip_k] f(p_k)=0 
  \quad \forall p_k=\frac{2 \pi}{L}k\;,\quad k\in\mathbb{Z}\;,
  \quad \text{with} \quad \n(p+\frac{2\pi}{a})=\n(p)\;.
  \label{eq:constraintder} \end{equation}
Eq.~(\ref{eq:constraintder}) can only be satisfied if for all  
momenta $p_k$ either the
Fourier component of the averaging function, $f(p_k)$, or the
difference between the continuum and the lattice derivative operator
vanishes. If $f(p_k)$ vanishes only outside the first Brillouin zone,
the only solution for $\n$ is the non-local SLAC-derivative
\cite{SLAC}, which can nevertheless be useful in some low dimensional models 
\cite{Kastner}.

If $f(p_k)$ also vanishes for momenta inside the first Brillouin zone,
other lattice derivative operators are allowed, but the averaging
function $f$ then introduces an effective cutoff below $2\pi/a$
\cite{Bergner:2008ws}. From a blocking perspective $f$ then generates
less independent blocked fields than present on the lattice. Note also
that the non-local derivative operator does not necessarily appear in
the action or the lattice symmetry operator, it is just included in
the naive lattice transformations $M$.

\sectionn{Free supersymmetry}
Now we have all ingredients to discuss supersymmetry on the lattice. 
For simplicity we consider supersymmetric quantum mechanics (SUSYQM).
For supersymmetric quantum mechanics the continuum multiplet $\vp$
consists of bosonic fields $\chi$ and $F$ (the latter being
non-dynamical) plus fermionic fields $\psi$ and $\bpsi$, which are
transformed into each other under supersymmetry.

The field content on the lattice and the naive lattice transformations 
\begin{equation}
\delta\left(
\begin{array}{c}
\chi \\ F \\ \psi \\ \bpsi
\end{array}\right)_n
=\left(\begin{array}{c c c c}
0&0&-\bveps&\veps\\
0&0&-\bveps\nabla&-\veps\nabla\\
-\veps\nabla&-\veps&0&0\\
\bveps\nabla&-\bveps&0&0
\end{array}\right)_{nm}
\left(
\begin{array}{c}
\vp \\ F \\ \psi \\ \bpsi
\end{array}\right)_m=(\veps \M+\bveps \bar{M})^{ij}_{nm} \phi^j_m\, ,
\end{equation}
are analogous to the continuum ones. The infinitesimal parameters
$\veps$ and $\bveps$ are Grassmannian and the SLAC-derivative, $\n$,
replaces the continuum derivative as discussed above. As a general
ansatz, the quadratic (off-shell) lattice action has the kernel
\begin{equation}
\label{eq:action_ansatz}
K^{ij}_{nm}=
a\left(
\begin{array}{c c c c}
-\Box_{nm}&-m_{b,nm}&0&0\\
-m_{b,nm}&-I_{nm}&0&0\\
0&0&0&(\tn-m_f)_{nm}\\
0&0&(\tn+m_f)_{nm}&0
\end{array}\right)\; .
\end{equation}
Translation invariance demands that all of these matrices are
circulant, i.e.\ they commute.  The matrices $I$, $\Box$, $m_{b,f}$
are symmetric, whereas $\tn$ is antisymmetric and in the continuum
limit they need to approach 1, $\partial^2$, $m$ and $\partial$,
respectively.
 
Together with the blocking kernel, these matrices are subject to our
relation (\ref{eq:matrix}). First we notice that a vanishing $\alm$,
or a symmetric one according to eq.~(\ref{eq:symmetric_alpha}), yields
a deformed symmetry equal to the naive one. These are non-local due to
the presence of the SLAC operator.

Hence we must use a nontrivial blocking kernel for generating local
lattice actions with local symmetries. First we choose $\alm$ diagonal
in lattice indices, as for the Ginsparg-Wilson relation, and -- due to
symmetry reasons -- symmetric in the bosonic sector and antisymmetric
in the fermionic sector. A solution of the coefficient matrices in the
action (\ref{eq:action_ansatz}) in terms of the blocking kernel and
the SLAC-derivative $\n$ is possible and non-trivial, see
\cite{Bergner:2008ws}. In particular, the lattice derivative
  $\tn$ in the action is proportional to $\n$ with a prefactor similar
  to a massive propagator $({\rm const.}^2-\n^2)^{-1}$. In the
  continuum similar expressions lead to an exponential decay for large
  distances. On the lattice, however, the corresponding behaviour is
  spoilt by terms decaying only algebraically, exactly because $\n$
  is finite at the boundary of the first Brillouin zone
  \cite{Bergner:2008ws}. 

  Since the blocking kernel $\alpha$ has no direct physical
  implication, we now adjust $\alm$ according to a local action and
  local deformed supersymmetry. With a general ansatz for $M_{\rm
    def}$ and after some computations we are left with a relation
  $\tn=I\n$ \cite{Bergner:2008ws}. As $\tn$ and $I$ appear in the
  lattice action and in the deformed symmetry, they must be local.
  This implies that $I$ vanishes together with all its derivative at
  the boundary of the Brillouin zone. In this delicate way a decay of
  the Dirac operator $\tn$ stronger than algebraic (but not
  exponential) can be achieved.
\sectionn{Interacting systems}
One of the important properties of our symmetry relation
(\ref{eq:relation}) is its validity for interacting theories.
Therefore, we can embark on constructing a local interacting supersymmetric
theory on the lattice in our approach. 

Beyond second order, however, the non-linearity of this relation starts
to play an important role. Since it connects different orders of
fields, solutions of the symmetry relation are generically
non-polynomial. This is not unexpected since the blocked action is
comparable to an effective action. In order to truncate the tower of
interactions, a necessary condition is (in matrix
notation)
\begin{equation}
\label{eq:Smax}
 \frac{\partial S^{\rm max}}{\partial \phi}\, M \alm 
\frac{\partial S^{\rm max}}{\partial \phi}=0\;,
\end{equation}
where $S^{\rm max}$ denotes the part of the action with the maximal
order in the fields. This means that this highest $S^{\rm max}$ does
not depend on the entire field multiplet. For SUSY, a purely bosonic
$S^{\rm max}$ suffices as $M$ mixes bosons with fermions. We have
illustrated this peculiar situation by virtue of an exact solution of
interacting SUSYQM restricted to constant fields, which depending on
the choice of $\alpha$ is either logarithmic or polynomial
\cite{Bergner:2008ws}, for chiral theories see \cite{Igarashi:2002ba}.
We stress that eq.~(\ref{eq:Smax}) represents only one relation of
those that follow from our relation
(\ref{eq:relation}) when ordered according to the powers of fields. 
\sectionn{Summary}
In this brief survey we have shown that symmetries of a continuum
action imply certain relations for the corresponding lattice action.
For general linear symmetries one can deduce a deformed symmetry
relation for the lattice theory \eq{eq:relation} that is a
generalisation of the standard GW-relation.

However the construction of lattice symmetry operators for space-time
dependent symmetries is subject to the obstruction \eq{eq:addconst}.
This constraint reflects the necessity to reconcile the averaging
procedure of the blocking with the space-time structure of the
symmetry. For space-time dependent symmetries it is non-trivial, which
applies in particular to derivative-dependent symmetries such as
supersymmetry, see \eq{eq:constraintder}, where we are led to the
non-local SLAC-derivative in the lattice symmetry operator.

For well-defined symmetry operators on the lattice we have to require
that both, the lattice operator $M_{\rm def}$ and the corresponding
lattice action are local. For non-local $M_{\rm def}$ and/or
lattice actions the fate of the symmetry in the continuum limit is
unclear as lattice artefacts might survive.

Within this setting we have discussed supersymmetric theories at the
example of supersymmetric quantum mechanics. We have shown that
solutions can be achieved in this case with a quadratic action as well
as in interacting theories. Although the symmetry relation
\eq{eq:relation} couples different orders of the fields, even for
interacting theories polynomial solutions might be possible. 

\vspace*{-0.1cm}
\newlength{\bibskip}
\setlength{\bibskip}{-0.2ex}


\begin{thebibliography}{99}
\vspace*{-0.1cm}
\bibitem{Dondi:1976tx}
  P.~H.~Dondi and H.~Nicolai,
  Nuovo Cim.\ {\bf A41} (1977) 1;
M.~Kato, M.~Sakamoto and H.~So,
JHEP {\bf 05} (2008) 057.
\bibitem{Nielsen}
  H.~B.~Nielsen and M.~Ninomiya,
  Phys.\ Lett.\  {\bf B105} (1981) 219;
   D.~Friedan,
     Commun.\ Math.\ Phys.\ {\bf 85} (1982) 481-490; 
  O.~Jahn and J.~M.~Pawlowski,
  Nucl.\ Phys.\  B {\bf 642} (2002) 357.
\bibitem{Ginsparg:1981bj}
  P.~H.~Ginsparg and K.~G.~Wilson,
  Phys.\ Rev.\  {\bf D25}, 2649 (1982).
\bibitem{Pawlowski:2005xe}
  J.~M.~Pawlowski,
  Annals Phys.\  {\bf 322} (2007) 2831.
\bibitem{Bergner:2008ws}
  G.~Bergner, F.~Bruckmann and J.~M.~Pawlowski,
  arXiv:0807.1110 [hep-lat].
\bibitem{So:1998ya}
  H.~So and N.~Ukita,
  Phys.\ Lett.\  {\bf B457} (1999) 314, 
 [arXiv:hep-lat/9812002].
\bibitem{Luscher}
  M.~L\"uscher,
  Phys.\ Lett.\  {\bf B428} (1998) 342.
\bibitem{SLAC}
     S.~D.~Drell, M.~Weinstein and S.~Yankielowicz,
     Phys.\ Rev.\ {\bf D14} (1976) 487;
     S.~D.~Drell, M.~Weinstein and S.~Yankielowicz,
     Phys.\ Rev.\ {\bf D14} (1976) 1627.
\bibitem{Kastner}
     T.~Kastner, G.~ Bergner, S.~Uhlmann, A.~Wipf and C.~Wozar,
     arXiv:0807.1905 [hep-lat].
\bibitem{Igarashi:2002ba}
  Y.~Igarashi, H.~So and N.~Ukita,
  Phys.\ Lett.\  {\bf B535} (2002) 363;
  Nucl.\ Phys.\  {\bf B640} (2002) 95. 

\end{thebibliography}
\end{document}